\documentclass[twocolumn,showpacs,superscriptaddress,aps,amsmath,amssymb]{revtex4}
\usepackage{amsfonts}
\usepackage{bm}

\newcommand{\ga}{\gamma}
\newcommand{\la}{\lambda}
\newcommand{\om}{\omega}
\newcommand{\Om}{\Omega}
\newcommand{\ve}{\varepsilon}
\newcommand{\dl}{\delta}
\newcommand{\bk}{{\bf k}}

\begin{document}
\title{Coherent quantum control of $\Lambda$-atoms through the stochastic limit}

\author{L. Accardi}
 \email{accardi@volterra.mat.uniroma2.it}
  \homepage[homepage ]{http://volterra.mat.uniroma2.it}
   \affiliation{Centro Vito Volterra, Universit\`a di Roma ``Tor Vergata'', 00133, Via Columbia 2, Roma, Italy}
\author{S. V. Kozyrev}
 \email{kozyrev@mi.ras.ru}
  \affiliation{Steklov Mathematical Institute, Russian
Academy of Sciences, Gubkin St. 8, 119991, Moscow, Russia}
\author{A. N. Pechen}
 \email{pechen@mi.ras.ru}
  \affiliation{Steklov Mathematical Institute, Russian
Academy of Sciences, Gubkin St. 8, 119991, Moscow, Russia}

\date{\today}

\begin{abstract}

We investigate, using the stochastic limit method, the coherent quantum control of a 3-level atom in
$\Lambda$-configuration interacting with two laser fields. We prove that, in the generic situation, this
interaction entangles the two lower energy levels of the atom into a single qubit, i.e. it drives at an
exponentially fast rate the atom to a stationary state which is a coherent superposition of the two lower levels.
By applying to the atom two laser fields with appropriately chosen intensities, one can create, in principle, any
superposition of the two levels. Thus {\it relaxation} is not necessarily synonymous of {\it decoherence}.
\end{abstract}

\maketitle

\section{Introduction}
Preparation of atoms and molecules in a predefined state plays an important role in modern atomic and molecular
physics, in particular in atom optics and quantum information. The difficulty of the problem consists in the fact
that, in order to control a system, one has to interact with it, but interaction introduces dissipative effects
and hence, at least generically, decoherence. In this note we prove, in specific but important and physically
realizable example, that in some cases dissipation can generate coherence.

One of the ways to drive the system, atom or molecule, to the desired state is to exploit its interaction with
laser pulses, i.e., to use the coherent laser control or laser-induced population transfer~\cite{a,vhsb}. In this
approach monochromatized and near-resonant with the atomic Bohr frequencies radiation fields are used to force the
system to the final state. The laser coherent control techniques can be used in applications to laser cooling
based on coherent population trapping~\cite{aak}, quantum computing with trapped ions~\cite{gzc,CiZo95}, etc.

The stochastic limit method~\cite{alv} was applied in~\cite{ak} to study the phenomenon of coherent population
trapping in an atom in lambda--configuration with a doubly degenerate ground state. It was found that the action
of the field drives the atom to a $1$--parameter family of stationary states so that the choice of one state in
this family is uniquely determined by the initial state of the atom and by the initial state of the field.

In~\cite{ak} the initial state of the field was chosen to be an arbitrary mean zero, gauge invariant Gaussian
state, e.g. the Fock vacuum or an equilibrium state at any temperature. On the other hand, in the usual
experiments on coherent population trapping, the atom is driven by two laser beams, i.e. coherent states, resonant
with the atomic frequencies and, up to now, there is no evidence of this $1$--parameter family of invariant
states~\cite{a}--\cite{sz}.

In the present paper we prove that, for a three-level non-degenerate atom, the field drives the atom to a pure
state, which is a superposition of the two lowest energy levels and that, applying to the atom two laser fields
with appropriately chosen intensities one obtains a single superposition. In the case of a degenerate atom we show
that, if one starts from a coherent state, then the family of invariant states is destroyed and the field drives
the atom to the dark state.

The results of the present paper, combined with those of~\cite{ak} also suggest that the emergence of the dark
state could be experimentally realized not only by tuning two lasers, but also in a generic equilibrium state, by
preparing the state of the atom in the domain of attraction of the dark state.

The present result is obtained as a consequence of a general property of the stochastic limit, which we find for
the first time in the paper, according to which the effect of replacing a mean zero, gauge invariant state of the
field by a coherent one, amounts to the addition of a hamiltonian term to the master equation while the
dissipative part of the generator remains the same [cf. (\ref{gen})--(\ref{hapag})].

This general property is proved in sections 2--4 of the present paper for an arbitrary atom. In sections 5 and 6
we specialize to the case of $2$- and 3-level atoms, find for these cases the explicit form of the stationary
atomic states [(\ref{ro2}) for a two-level, (\ref{ro3.1}) and~(\ref{statpsi}) for a three-level atom], and prove
the results mentioned in the present introduction. The master equation we find in the 3-level case coincides with
the optical Bloch equation considered by Arimondo in~\cite{a} in the absence of scattering [Eq.~(2.7)] with the
only difference that we obtain the equation and the explicit formula for the coefficients $\Om_j$ [cf.
equations~(\ref{com}) and~(\ref{capom}) below] without assuming that the field is classical. Instead, we start
from the microscopic quantum dynamics of the total atom+radiation system.

The approach of the present paper to study the coherent quantum control or laser induced population transfer is
based on the derivation, using the stochastic limit method, of a quantum white noise equation approximating the
total dynamics in the weak coupling regime (Eq.~(\ref{wneq}), for a rigorous treatment see~\cite{alv}). The
programme to exploit exponentially fast decoherence as a control tool to drive an atom into a pre-assigned state
was formulated in~\cite{ai,ao}, where the control parameter is the interaction Hamiltonian. In the approach of the
present paper the control parameter is the state of the radiation, which can be easily controlled in experiments
and, instead of decoherence, it uses relaxation, due to dissipative dynamics, to create quantum coherence between
the atomic states.

\section{An atom in a laser field}

The dynamics of an atom interacting with radiation is determined on the microscopic level by the interaction
Hamiltonian and by the reference state of radiation.

The Hamiltonian describing an atom interacting with radiation is given by the sum of the free and interaction
terms
\[
H_\la=H_{\rm free}+H_{\rm int}=H_{\rm A}\otimes1+1\otimes H_{\rm R}+\la H_{\rm int},
\]
where $H_{\rm A}$ and $H_{\rm R}$ are free Hamiltonians of the atom and radiation, $H_{\rm int}$ interaction
Hamiltonian, and $\la$ is the coupling constant. The free Hamiltonian of the atom has discrete spectrum
\[
H_{\rm A}=\sum\limits_n\ve_n P_n,
\]
where $\ve_n$ is an eigenvalue and $P_n$ is the corresponding projection. The free Hamiltonian of the radiation is
\[
H_{\rm R}=\int d\bk\om(\bk)a^+(\bk)a(\bk),\qquad\om(\bk)=|\bk|,
\]
where the creation and annihilation operators $a^+(\bk)$, $a(\bk)$ describe creation and annihilation of photons
with momentum $\bk$. The interaction Hamiltonian has the form
\[
H_{\rm int}=i(D\otimes a^+(g)-D^+\otimes a(g)),
\]
where the operators $D$ and $D^+$ describe the transitions between the atomic levels, $a^+(g)=\int d\bk
g(\bk)a^+(\bk)$ is the smeared creation operator, and the formfactor $g(\bk)$ describes the coupling of the atom
with the $\bk$ mode of the field.

The state of the radiation, which corresponds to a long time laser pulse of frequency $\om$ and intensity $f$ is a
coherent state, which is determined by the coherent vector
\[
\tilde\Psi_\la=W\left(\la\int\limits_{S/\la^2}^{T/\la^2}S_te^{-it\om}fdt\right)\Phi_0,
\]
where $\Phi_0$ is the vacuum, $W(\cdot)$ the Weyl operator, which is defined as $W(f)=\exp[i(a(f)+a^+(f))]$,
$S_t=e^{it\om(\bk)}$ the one photon free evolution, and $[S/\la^2,T/\la^2]$ the time interval in which the laser
pulse is active. This vector $\tilde\Psi_\la$ is an eigenvector of the annihilation operator:
\[
a(g)\tilde\Psi_\la=c_\la\tilde\Psi_\la.
\]
The eigenvalue
\[
c_\la=\la\int\limits_{S/\la^2}^{T/\la^2}dt\int d\bk g^*(\bk)f(\bk)e^{it(\om(\bk)-\om)}
\]
is determined by the intensity of the laser $f$ and the form factor $g$. The pulse starts at time $S/\la^2$ and
ends at time $T/\la^2$. Since the coupling constant $\la$ is small the duration of the pulse, being of order
$\la^{-2}$, is large. The function $f(\bk)$ describes the amplitude of the input field with momentum $\bk$.

The state, which corresponds to several laser pulses of frequencies $\om_1,\dots,\om_n$, intensities
$f_1,\dots,f_n$, and acting during time intervals $[S_1,T_1],\dots,[S_n,T_n]$, is determined by the coherent
vector
\begin{equation}\label{psi}
\Psi_\la=W\left(\la\sum\limits_{l=1}^n\int\limits_{S_l/\la^2}^{T_l/\la^2}S_te^{-it\om_l}f_ldt\right)\Phi_0.
\end{equation}

The initial state of the total system is supposed to be factorized
\[
\om=\om_{\hat \rho}\otimes\om_\la={\rm Tr}\,(\hat\rho\,\cdot)\otimes\om_\la,
\]
where $\om_{\hat \rho}$ is the state of the atom determined by a density matrix $\hat\rho$ and $\om_\la$ is the
state of the radiation determined by the coherent vector $\Psi_\la$:
$\om_\la=\langle\Psi_\la,\cdot\Psi_\la\rangle$.

The dynamics of the total system is determined by the evolution operator which in interaction picture is defined
as
\[
U_\la(t)=e^{itH_{\rm free}}e^{-itH_\la},
\]
where the index denotes the dependence on the coupling constant. The evolution operator determines the Heisenberg
evolution of any observable of the atom, say $X$, as
\[
X_\la(t)=U^+_\la(t)XU_\la(t).
\]
The reduced dynamics of the observable is obtained by taking average over the state of the radiation
\[
\langle X_\la(t)\rangle:=\langle\Psi_\la,U^+_\la(t)XU_\la(t)\Psi_\la\rangle
\]
and, by duality, it can be equivalently described in terms of the reduced density matrix $\hat\rho_\la(t)$,
defined by the following equality:
\begin{eqnarray*}
{\rm Tr}_{\rm A}(\hat\rho_\la(t)X)={\rm Tr}_{\rm A}(\hat\rho\langle X_\la(t)\rangle) \\
={\rm Tr}_{\rm A}\Bigl\{\overline{\rm Tr}_{\rm
R}\Bigl[U_\la(t)\Bigl(\hat\rho\otimes|\Psi_\la\rangle\langle\Psi_\la|\Bigr)U^+_\la(t)\Bigr]X\Bigr\},
\end{eqnarray*}
where ${\rm Tr}_{\rm A}$ denotes the trace over the atomic degrees of freedom and $\overline{\rm Tr}_{\rm R}$ the
partial trace over the field degrees of freedom.

The first step of the present paper is to give a non fenomenological derivation, from the first principles, of the
quantum master equation for the reduced density matrix of the atom using the stochastic limit method. This is done
in the following two sections.

\section{The stochastic limit}
It is impossible to find, for a general coupling $D$, the explicit form of the dynamics for the reduced density
matrix for realistic models, while the weakness of the interaction suggests some approximations. For small time
the dynamics can be effectively studied by using the perturbation series. However, the $n$-th term of the series
behaves like $(\la^2 t)^n$ and the approximation with the lowest order terms of the series becomes invalid for
time which is large enough.

Hence the study of the long time dynamics requires another approach. Such an approach to study the dynamics on the
long time scale ($\sim\la^{-2}$) is the stochastic limit method~\cite{alv}. In the stochastic limit one considers
the long time dynamics (on the time scale of order $\la^{-2}$) for a system with weak interaction. Mathematically
this means that one takes the limit as the coupling constant goes to zero, $\la\to 0$, time goes to infinity,
$t\to\infty$, but in such a way that the quantity $\la^2t$ remains fixed. In this limit the dynamics of the total
system is given by the solution of a quantum white noise or stochastic differential equation (which is a unitary
adapted process). Since the interaction of an atom with radiation is weak one can apply the stochastic limit
procedure to study the long time dynamics of this system.

On the long time scale the behavior of the exact reduced density matrix is approximated by the limiting density
matrix
\[
\hat\rho(t)=\lim\limits_{\la\to 0}\hat\rho_\la(t/\la^2)
\]
so that $\rho_\la(t)\simeq\rho(\la^2t)$. The quantum master equation for the limiting density matrix in the case
when the radiation is in a Gibbs state was derived and discussed in many papers (see~\cite{alv} and references
therein). The purpose of the present paper is to derive the quantum master equation for the limiting density
matrix $\hat\rho(t)$ in the case the radiation is in a coherent state and to study this equation for particular,
but important cases of two and three-level atoms.

The evolution operator after the time rescaling satisfies the equation
\begin{eqnarray*}
\frac{dU_\la(t/\la^2)}{dt}\\
=\sum\limits_{\om\in B}\left(D_\om\otimes a^+_{\la,\om}(t)-D^+_\om\otimes a_{\la,\om}(t)\right)U_\la(t/\la^2),
\end{eqnarray*}
where $B$ is the set of all Bohr frequencies of the atom (spectrum of the free atomic Liouvillian $i[H_{\rm
A},\cdot]$, i.e. the set of $\om=\ve_n-\ve_m$, where $\ve_n,\ve_m$ are eigenvalues of the free atomic
Hamiltonian), for each $\om\in B$
\begin{equation}\label{resca}
a_{\la,\om}(t)=\frac{1}{\la}\int d\bk g^*(\bk)e^{-i(\om(\bk)-\om)t/\la^2}a(\bk)
\end{equation}
is the rescaled time dependent annihilation operator, and the operator
\begin{equation}\label{dom}
D_\om=\sum\limits_{n,m:\, \varepsilon_n-\varepsilon_m=\om}P_mDP_n
\end{equation}
describes transitions between atom levels with energies $\ve_n$ and $\ve_m$ with the energy difference
$\ve_n-\ve_m=\om$.

In the stochastic limit one first proves that the time rescaled creation and annihilation operators converge, as
$\la\to 0$, to a quantum white noise (for details of the stochastic limit procedure see~\cite{alv}):
\[
\lim\limits_{\la\to 0} a_{\la,\om}(t)= b_\om(t),
\]
where the quantum white noise operators $b_\om(t)$ are $\dl$-correlated in time and satisfy the commutation
relations
\begin{equation}\label{comm}
[b_\om(t),b^+_{\om'}(t')]=\dl(t'-t)\dl_{\om,\om'}2{\rm Re}\,\ga_\om.
\end{equation}
The complex number
\begin{equation}\label{sus}
\ga_\om=\int d\bk\frac{|g(\bk)|^2}{i(\om(\bk)-\om-i0)}
\end{equation}
is the generalized susceptibility and its real part
\[
{\rm Re}\,\ga_\om=\pi\int d\bk|g(\bk)|^2\dl(\om(\bk)-\om)
\]
gives the decay rate of the $\om$-transition (cf.~\cite{alv}, sect. 4.20). The Kronecker $\dl$-symbol
$\dl_{\om,\om'}$ in~(\ref{comm}) indicates the mutual independence of the white noise operators for different Bohr
frequencies.

This allows us to derive the equation for the limiting evolution operator
\[
U_t:=\lim\limits_{\la\to 0}U_\la(t/\la^2).
\]
That is the white noise Schr\"odinger equation
\begin{equation}\label{wneq}
\frac{dU_t}{dt}=-iV(t)U_t
\end{equation}
with the white noise Hamiltonian
\[
V(t)=i\sum\limits_{\om\in B}\Bigl(D_\om\otimes b^+_\om(t)-D^+_\om\otimes b_\om(t)\Bigr).
\]

Let $X$ be any observable of the atom and $X_t=U^+_tXU_t\equiv\lim_{\la\to 0}X_\la(t/\la^2)$ its limiting time
evolution. Using the stochastic golden rule of the stochastic limit (cf.~\cite{alv}, sect. 5.9), one gets the
quantum Langevin equation for $X_t$, that is a normally ordered white noise differential equation:
\begin{eqnarray}
\frac{dX_t}{dt}=U^+_t\Theta(X)U_t\nonumber\\
+\sum\limits_\om\Bigl(b^+_\om(t) U^+_tL^+_\om(X)U_t+U^+_tL_\om(X)U_tb_\om(t)\Bigr)\label{qL}
\end{eqnarray}
(such an equation is equivalent to a certain quantum stochastic differential equation). Here
\[
\Theta(X)=\sum\limits_\om\Bigl(2{\rm Re}\,\ga_\om D^+_\om XD_\om-\ga_\om XD^+_\om D_\om- \bar\ga_\om D^+_\om D_\om
X\Bigr)
\]
is the generator of a quantum Markov semigroup and, in the notation~(\ref{dom}),
\[
L^+_\om(X)=[X,D_\om],\qquad L_\om(X)=[D^+_\om,X].
\]
The normally ordered form of the quantum Langevin equation, when the annihilation white noise operators are put on
the right of the evolution operator and creation operators are put on the left, is very convenient for study the
reduced dynamics in the case the radiation is in a coherent state. In order to get the reduced dynamics one just
has to average equation~(\ref{qL}) over the state of radiation.

The coherent state $\Psi_\la$ of the radiation is approximated by the limiting coherent vector
\[
\Psi=\lim\limits_{\la\to 0}\Psi_\la.
\]
This vector is an eigenvector of the quantum white noise annihilation operator:
\begin{equation}\label{eq1}
b_{\om_l}(t)\Psi=\chi_{[S_l,T_l]}(t)c_{\om_l}\Psi.
\end{equation}
Here $\chi_{[S_l,T_l]}(t)$ is the characteristic function of the interval $[S_l,T_l]$, determining the duration of
the pulse. The complex number $c_{\om_l}$ depends on the state of the radiation as follows:
\begin{equation}\label{com}
c_{\om_l}=2\pi\int d\bk g^*(\bk)f_l(\bk)\dl(\om(\bk)-\om_l)
\end{equation}
and determines the Rabi frequencies of the field.

In fact, the coherent vector $\Psi_\la$ is an eigenvector of the time rescaled annihilation operator
$a_{\la,\om_l}(t)$ (defined by~(\ref{resca})) with eigenvalue
\begin{eqnarray*}
\sum\limits_{m=1}^n\int\limits_{S_m}^{T_m}d\tau\frac{1}{\la^2}\int d\bk
g^*(\bk)f_l(\bk)e^{i(\tau-t)(\om(\bk)-\om_l)/\la^2}\\
\times e^{i\tau(\om_l-\om_m)/\la^2}.
\end{eqnarray*}
From that, using the limit (see~\cite{alv} for details)
\begin{eqnarray*}
\lim\limits_{\la\to 0}\frac{1}{\la^2}e^{i(\tau-t)(\om(\bk)-\om_l)/\la^2} e^{i\tau(\om_l-\om_m)/\la^2}\\
=2\pi\dl(t-\tau)\dl_{\om_l,\om_m}\dl(\om(\bk)-\om_l)
\end{eqnarray*}
one obtains the limiting eigenvalue
\[
2\pi\sum\limits_{m=1}^n\int\limits_{S_m}^{T_m}d\tau\dl(t-\tau)\int d\bk
g^*(\bk)f_m(\bk)\dl_{\om_l,\om_m}=\chi_{[S_l,T_l]}(t)c_{\om_l}
\]
We will consider the case $S_l=S$, $T_l=T$ for all $l$, which means that all lasers are applied during the same
time interval $[S,T]$.

Let $X$ be an observable of the atom and denote by
\[
\langle X\rangle_t= \langle\Psi,X_t\Psi\rangle
\]
the average, over the limiting state of the radiation, of its time evolution. Taking the average of both sides of
the quantum Langevin equation~(\ref{qL}) and using property~(\ref{eq1}) one gets the following equation for the
averaged observable:
\begin{eqnarray}
\frac{d\langle X\rangle_t}{dt}=\langle \Theta(X)\rangle_t\nonumber\\
+i\chi_{[S,T]}(t)\sum\limits_\om\langle c_\om L_\om(X)-c^*_\om L^+_\om(X)\rangle_t.\label{eqX}
\end{eqnarray}
This master equation for the case with presence of external field is one of the main results of the present paper.
In the next section we will rewrite it in the equivalent form as an equation for the reduced density matrix.

\section{The quantum master equation}

The dynamics of the atom can be described either in terms of observables, or equivalently, in terms of density
matrices. The equation for the reduced density matrix can be obtained using the relation
\[
{\rm Tr}(\hat\rho(t)X)={\rm Tr}(\hat\rho(0)\langle X\rangle_t)
\]
and equation~(\ref{eqX}). It is a quantum master equation of the form:
\begin{equation}\label{masteqro}
\frac{d\hat\rho(t)}{dt}={\cal L}_t(\hat\rho(t)).
\end{equation}
The time dependent generator of this equation is given by the sum of its dissipative and Hamiltonian parts:
\begin{equation}
{\cal L}_t(\hat\rho)={\cal L}_{\rm diss}(\hat\rho)-i[H_{\rm eff}(t),\hat\rho].\label{gen}
\end{equation}
As already stated in the Introduction, the dissipative part is the same as in the case the reservoir in the vacuum
state:
\begin{equation}
{\cal L}_{\rm diss}(\hat\rho)=\sum\limits_\om {\rm Re}\,\ga_\om\Bigl(2D_\om\hat\rho D^+_\om-\hat\rho D^+_\om
D_\om-D^+_\om D_\om\hat\rho \Bigr).\label{dipag}
\end{equation}
The Hamiltonian part is determined by the effective time dependent Hamiltonian
\begin{equation}
H_{\rm eff}(t)=\sum\limits_\om\Bigl({\rm Im}\ga_\om D^+_\om D_\om+\chi_{[S,T]}(t)(c^*_\om D_\om+c_\om
D^+_\om)\Bigr).\label{hapag}
\end{equation}
The first term in the brackets is the standard term which appears in the stochastic limit when the reservoir is in
the vacuum state. It commutes with the free atomic Hamiltonian $H_{\rm A}$. The other terms are new and their
presence leads to the important consequence that the effective Hamiltonian at time $t\in[S,T]$ does not commute
with the free atom Hamiltonian
\begin{equation}\label{nc}
[H_{\rm eff}(t),H_{\rm A}]\ne 0.
\end{equation}

Therefore the states of the atom driven by a long time laser pulse, which are approximated by stationary states of
the master equation obtained from~(\ref{masteqro}) replacing in the effective Hamiltonian~(\ref{hapag})
$\chi_{[S,T]}(t)$ by identity, will not be diagonal with respect to $H_{\rm A}$. Our goal is to investigate the
structure of these states. In generic situation the interaction with the laser field drives the atom to a
(dynamical) equilibrium state at an exponential rate. In this sense the interaction prepares a state of the atom.
Thus our problem can be reformulated as follows: What kind of states can we prepare applying a laser pulse? For a
long laser pulse these states are described by the time independent generator ${\cal L}$ obtained
from~(\ref{eqX}), by setting $S\to-\infty$, $T\to+\infty$, i.e. the characteristic function $\chi_{[S,T]}$
in~(\ref{eqX}) replaced by the identity.

In the limit $S\to-\infty$, $T\to+\infty$ the effective Hamiltonian becomes time independent
\[
H_{\rm eff}=\sum\limits_\om\Bigl({\rm Im}\ga_\om D^+_\om D_\om+c^*_\om D_\om+c_\om D^+_\om\Bigr).
\]
In this case the stationary state $\hat\rho_{\rm st}$ is a solution of the following equation
\begin{equation}\label{st}
{\cal L}(\hat\rho_{\rm st})\equiv{\cal L}_{\rm diss}(\hat\rho_{\rm st})-i[H_{\rm eff},\hat\rho_{\rm st}]=0.
\end{equation}
In the following sections we solve explicitly this equation for two and three-level atoms.

\section{Stationary states for a two-level atom}
Let us consider a two-level atom under the rotating wave approximation.

The free Hamiltonian of the atom is
\[
H_{\rm A}=\ve_0|0\rangle\langle 0|+\ve_1|1\rangle\langle 1|,
\]
where $\ve_0$ and $\ve_1$ are the energies of the ground and the excited states, so that the Bohr frequency
$\om=\ve_1-\ve_0>0$ is positive. The ground state is denoted by $|0\rangle$ and the excited state by $|1\rangle$.
The transitions between atomic levels are described by the operators
\[
D=|0\rangle \langle 1|\equiv\sigma_-,\qquad D^+=|1\rangle \langle 0|\equiv\sigma_+.
\]
The interaction Hamiltonian has the form
\[
H_{\rm int}=i(\sigma_-\otimes a^+(g)-\sigma_+\otimes a(g)).
\]
The first term in the interaction Hamiltonian describes the process of emission of a photon due to transition of
the atom from the excited to the ground state. The second term describes the opposite process, that is the
absorbtion of a photon by the atom and transition to the excited state.

The real part  $\ga={\rm Re}\,\ga_\om$ of the susceptibility~(\ref{sus}) determines the inverse lifetime of the
atom without interaction with external field and the imaginary part $\Delta\om={\rm Im}\ga_\om$ determines energy
shift (cf.~\cite{alv}, sect. 5.5). The dissipative part of the generator and the effective Hamiltonian for this
system have the forms
\begin{eqnarray*}
{\cal L}_{\rm diss}(\hat\rho)=\ga(2\sigma_-\rho\sigma_+-\rho\sigma_+\sigma_--\sigma_+\sigma_-\rho),\\
H_{\rm eff}=\Delta\om|1\rangle\langle 1|+\Om^*\sigma_-+\Om\sigma_+,
\end{eqnarray*}
where $\Om\equiv c_\om$ is the complex Rabi frequency of the laser field. Equation~(\ref{st}) for a stationary
state $\hat\rho_{\rm st}$ of the two-level atom equivalent to the following equations on the matrix elements
$\rho_{mn}=\langle m|\hat\rho_{\rm st}|n\rangle$:
$$
\begin{array}{l}
\ga\rho_{11}={\rm Im}\,(\Om\rho_{01})\nonumber\\
\ga_\om\rho_{10}=i\Om(\rho_{11}-\rho_{00}).\nonumber
\end{array}
$$
The general properties ${\rm tr}\hat\rho=1$, $\hat\rho^+=\hat\rho$ of the density matrix lead to the relations
$\rho_{00}+\rho_{11}=1$, $\rho_{01}=\rho^*_{10}$, $\rho^*_{11}=\rho_{11}$. Using these relations one finds that,
with the notation $\alpha=-i\Om/\ga_\om$, the stationary state is
\begin{equation}\label{ro2}
\hat\rho_{\rm st}\equiv\left(\begin{array}{cc}
\rho_{11} & \rho_{10}\\
   & \\
\rho_{01} & \rho_{00}
 \end{array}
 \right)
=\frac{1}{1+2|\alpha|^2}\left(
 \begin{array}{cc}
   |\alpha|^2& \alpha\\
   & \\
  \alpha^*& 1+|\alpha|^2
 \end{array}\right)
\end{equation}
This state is pure only if $\alpha=0$. For very intense illumination ($|\Om|/|\ga_\om|\gg 1$) the quantity
$|\alpha|$ becomes very large and the atom becomes saturated with equal probabilities in the upper and lower
levels, so that $\rho_{00}=\rho_{11}=1/2$, $\rho_{01}=0$. For very low illumination, the stationary state is the
ground state.

\section{Stationary states for a three-level atom}

The free Hamiltonian of a three-level atom is
\[
H_{\rm A}=\ve_0|0\rangle\langle 0|+\ve_1|1\rangle\langle 1|+\ve_2|2\rangle\langle 2|,
\]
where $|0\rangle$, $|1\rangle$ and $|2\rangle$ denote the ground state, the first and second excited states. For a
non degenerate atom the energies satisfy the inequalities $\ve_2>\ve_1>\ve_0$. There are three positive Bohr
frequencies $\om_1=\ve_1-\ve_0$, $\om_2=\ve_2-\ve_1$, $\om_3=\ve_2-\ve_0$, which correspond to three possible
transitions. The transitions between the atomic levels due to interaction with radiation are described by the
operator
\[
D=d^*_1|0\rangle\langle 1|+d^*_2|1\rangle\langle 2|+d^*_3|0\rangle\langle 2|,
\]
where $d_j$ are some complex numbers. In particular, the contribution of the first term in the sum to the
interaction Hamiltonian corresponds to emission of a photon by the atom and transition from the first excited
level $|1\rangle$ to the ground state $|0\rangle$.

For the three-level atom the three possible transitions are described by the operators
\begin{eqnarray}
D_1:=D_{\om_1}=d^*_1|0\rangle\langle 1|,\quad & D_2:=D_{\om_2}=d^*_2|1\rangle\langle 2|,\nonumber\\
& D_3:=D_{\om_3}=d^*_3|0\rangle\langle 2|. \label{dfdj}
\end{eqnarray}
Hence, with the notations
\begin{equation}
\ga_j:=\ga_{\om_j}|d_j|^2=a_j+ib_j,\label{dfajbj}
\end{equation}
($a_j$ and $b_j$ are the real and imaginary parts of $\ga_j$) the dissipative part of the generator, in the
generic case, can be written as
\begin{eqnarray*}
{\cal L}_{\rm diss}(\hat\rho)=2\left(a_1\rho_{11}+a_3\rho_{22}\right)|0\rangle\langle
0|+2a_2\rho_{22}|1\rangle\langle 1|\\
\vphantom{\int}-a_1(\hat\rho|1\rangle\langle 1|+|1\rangle\langle 1|\hat\rho)-(a_2+a_3)(\hat\rho|2\rangle\langle
2|+|2\rangle\langle 2|\hat\rho),
\end{eqnarray*}
where $\rho_{mn}=\langle m|\hat\rho|n\rangle$ are the matrix elements of the density matrix. Introducing the Rabi
frequencies of the laser field
\begin{equation}\label{capom}
\Om_j:=c_{\om_j}d_j,
\end{equation}
the effective Hamiltonian can be written as
\begin{eqnarray*}
H_{\rm eff}=b_1|1\rangle\langle 1|+(b_2+b_3)|2\rangle\langle 2|\\
+\Om_1|1\rangle\langle 0|+\Om_2|2\rangle\langle 1|+\Om_3|2\rangle\langle 0|\\
+\Om^*_1|0\rangle\langle 1|+\Om^*_2|1\rangle\langle 2|+\Om^*_3|0\rangle\langle 2|.
\end{eqnarray*}
The matrix elements of the stationary state for arbitrary Rabi frequencies $\Om_j$ satisfy the following system of
equations:
$$
\begin{array}{l}
\vphantom{\int\limits_0}  a_1\rho_{11}+a_3\rho_{22}={\rm Im}\,(\Om_1\rho_{01}+\Om_3\rho_{02})\\
\vphantom{\int\limits_0} (a_2+a_3)\rho_{22}={\rm Im}\,(\Om_2\rho_{12}+\Om_3\rho_{02})\\
\vphantom{\int\limits_0} (\ga^*_2+\ga^*_3)\rho_{02}+i\Om^*_3(\rho_{22}-\rho_{00})-i\Om^*_2\rho_{01}+i\Om^*_1\rho_{12}=0\\
\vphantom{\int\limits_0} \ga^*_1\rho_{01}+i\Om^*_1(\rho_{11}-\rho_{00})-i\Om_2\rho_{02}+i\Om^*_3\rho_{21}=0\\
\vphantom{\int\limits_0}
(\ga_1+\ga^*_2+\ga^*_3)\rho_{12}+i\Om^*_2(\rho_{22}-\rho_{11})+i\Om_1\rho_{02}=i\Om^*_3\rho_{10}.
\end{array}
$$

Let us consider the case when the radiation, describing laser pulse, is in a coherent state  which drives only
transitions from the ground state to the highest excited state, i.e., a laser field with frequency near $\om_3$.
In this case $\Om_1=\Om_2=0$, $\Om_3\equiv \Om\ne 0$ is the only nonzero Rabi frequency and the system of
equations for the matrix elements of the stationary state becomes simpler:
$$
\begin{array}{l}
\vphantom{\int\limits_0} a_1\rho_{11}+a_3\rho_{22}={\rm Im}\,(\Om\rho_{02})\\
\vphantom{\int\limits_0} (a_2+a_3)\rho_{22}={\rm Im}\,(\Om\rho_{02})\\
\vphantom{\int\limits_0} (\ga^*_2+\ga^*_3)\rho_{02}+i\Om^*(\rho_{22}-\rho_{00})=0\\
\vphantom{\int\limits_0} \ga^*_1\rho_{01}+i\Om^*\rho_{21}=0\\
\vphantom{\int\limits_0} (\ga_1+\ga^*_2+\ga^*_3)\rho_{12}-i\Om^*\rho_{10}=0.
\end{array}
$$
This system of equations, together with the conditions ${\rm tr}\,\hat\rho=1$, $\hat\rho^+=\hat\rho$, can be
easily solved. With the notations
\[
\alpha=\frac{-i\Om}{\ga_2+\ga_3}, \qquad r=\frac{a_2}{a_1}
\]
the solution, which is the stationary state, has the form
\begin{eqnarray}
\hat\rho_{\rm st}\equiv\left(\begin{array}{ccc}
\rho_{22} & \rho_{21} & \rho_{20} \\
   & & \\
\rho_{12} & \rho_{11} & \rho_{10} \\
   & & \\
\rho_{02} & \rho_{01} & \rho_{00}
 \end{array}
 \right)\nonumber\\
=\frac{1}{1+(2+r)|\alpha|^2}\left(
 \begin{array}{ccc}
  |\alpha|^2 & 0 & \alpha\\
   & & \\
  0 & r|\alpha|^2 & 0\\
   & & \\
  \alpha^* & 0 & 1+|\alpha|^2
\end{array}\right).\label{ro3.1}
\end{eqnarray}
The obtained density matrix $\hat\rho_{st}$ is the stationary state of the three-level atom in the laser field
driving only transitions from the ground state to the highest energy state of the atom.

If the ratio $r$ is very large, i.e. $a_2\gg a_1$, then the stationary state describes the population inversion
between the first two excited atomic levels. In this case the population of the first excited state of the atom is
greater than that of the ground state. To find the physical conditions for this effect let us note that the
quantity
\[
a_1=|d_1|^2\int d\bk|g(\bk)|^2\dl(\om(\bk)-\om_1)
\]
determines the inverse lifetime of the atom at the level $|1\rangle$ in the free space. That is $\tau\sim 1/a_1$.
Hence, if the lifetime $\tau$ of the atom in the first excited state is long enough, then we obtain the population
inversion between the ground and the first excited level.

As an example, consider a three-level atom in $\Lambda$ configuration, so that the transitions between the ground
state $|0\rangle$ and the first excited state $|1\rangle$ are forbidden. This means that $d_1=0$. In this case
$a_1=0$, the lifetime of the first excited level is infinite, and the steady state is $|1\rangle\langle 1|$, i.e.,
in this case all the population is concentrated on the first excited state.

\section{Three-level lambda-atom}

Let us now consider a three-level atom in $\Lambda$ configuration in the general coherent field, that is both Rabi
frequencies $\Om_2$ and $\Om_3$~(\ref{capom}) are different from zero.

The system of dynamical equations for matrix elements of the density matrix is (matrix elements are time
dependent, $\rho_{nm}\equiv\rho_{nm}(t)$, the dot denotes time derivative)
\[
\begin{array}{l}
\vphantom{\int\limits_0} \dot{\rho}_{00}=2a_3\rho_{22}-2{\rm Im}\,(\Om_3\rho_{02})\\
\vphantom{\int\limits_0} \dot{\rho}_{22}=-2(a_2+a_3)\rho_{22}+2{\rm Im}\,(\Om_2\rho_{12}+\Om_3\rho_{02})\\
\vphantom{\int\limits_0} \dot{\rho}_{02}=-(\ga^*_2+\ga^*_3)\rho_{02}-i\Om^*_3(\rho_{22}-\rho_{00})+i\Om^*_2\rho_{01}\\
\vphantom{\int\limits_0} \dot{\rho}_{01}=i\Om_2\rho_{02}-i\Om^*_3\rho_{21}\\
\vphantom{\int\limits_0}
\dot{\rho}_{12}=-(\ga^*_2+\ga^*_3)\rho_{12}-i\Om^*_2(\rho_{22}-\rho_{11})+i\Om^*_3\rho_{10}.
\end{array}
\]
The stationary solution of this system of equations is
\begin{eqnarray*}
\hat\rho_{\rm st}\equiv\left(\begin{array}{ccc}
\rho_{22} & \rho_{21} & \rho_{20} \\
   & & \\
\rho_{12} & \rho_{11} & \rho_{10} \\
   & & \\
\rho_{02} & \rho_{01} & \rho_{00}
 \end{array}
 \right)\\
=\frac{1}{|\Om_2|^2+|\Om_3|^2}\left(
 \begin{array}{ccc}
  0 & 0 & 0\\
   & & \\
  0 & |\Om_2|^2 & -\Om^*_2\Om_3\\
   & & \\
  0 & -\Om_2\Om^*_3 & |\Om_3|^2
\end{array}\right).
\end{eqnarray*}
An important property is that it is a pure state: $\hat\rho_{\rm st}=|\psi\rangle\langle\psi|$, where
\begin{equation}\label{statpsi}
|\psi\rangle=\frac{1}{\sqrt{|\Om_2|^2+|\Om_3|^2}}\Bigl(\Om^*_2|1\rangle-\Om^*_3|0\rangle\Bigr).
\end{equation}
In particular, if the ground state is doubly degenerate and if the coupling does not distinguish between the two
ground states, then $\Om_2=\Om_3$ and the unique stationary state is the dark state (cf.~\cite{a} for a
discussion).

\section{Conclusions}
In the present work we derive, starting from exact microscopic dynamics and using the stochastic limit method, the
quantum master equation~(\ref{masteqro}) for an arbitrary atom driven by a general laser field. As an application
of this equation, we study the cases of two and three-level atoms. The explicit form of the stationary states of
such atoms are found. It is shown that for a three-level lambda-atom in two laser fields, which drive both allowed
transitions, the stationary state of the atom is a pure state given by a superposition of the two lowest energy
levels. Varying the intensities of the lasers, any such a superposition can be obtained. Hence the two lower
levels of the atom behave like a single qubit and any state of this qubit can be obtained in a stable way and in
exponentially small time.

\begin{acknowledgments}
This work is supported by the EU Network QP-Applications, Contract No. HPRN-CT-2002-00279. A.Pechen and S. Kozyrev
are grateful to Luigi Accardi for kind hospitality in the Centro Vito Volterra. A. Pechen and S. Kozyrev are
partially supported by CNR-NATO fellowship and the grant of Russian Foundation for Basic Research RFFI
02-01-01084. S. Kozyrev is also partially supported by the grants CRDF UM1-2421-KV-02 and a Grant for support of
scientific schools NSh 1542.2003.1.

\end{acknowledgments}

\end{document}